\documentclass[aps,prl,superscriptaddress]{revtex4-1}
\usepackage{geometry}
\usepackage{setspace}
\usepackage{times}
\usepackage{fullpage}
\usepackage{color}
\usepackage{graphicx}
\begin{document}

\title{High-field transport properties of a P-doped BaFe$_2$As$_2$ film on technical substrate}

\author{Kazumasa\,Iida}
\email[Correspondence to:\,]{iida@nuap.nagoya-u.ac.jp}
\affiliation{Department of Crystalline Materials Science, Nagoya University, Chikusa-ku, Nagoya 464-8603, Japan}
\author{Hikaru\,Sato}
%\email[Electronical address:\,]{h_sato@lucid.msl.titech.ac.jp}
\affiliation{Laboratory for Materials and Structures, Institute of Innovative Research, Tokyo Institute of Technology, Mailbox R3-1, 4259 Nagatsuta-cho, Midori-ku, Yokohama 226-8503, Japan}
\author{Chiara\,Tarantini}
%\email[Correspondence to:\,]{tarantini@asc.magnet.fsu.edu}
\affiliation{Applied Superconductivity Center, National High Magnetic Field Laboratory, Florida State University, Tallahassee FL 32310, USA}
\author{Jens\,H\"{a}nisch}
%\email[Correspondence to:\,]{jens.haenisch@kit.edu}
\affiliation{Karlsruhe Institute of Technology, Institute for Technical Physics, Hermann-von-Helmholtz-Platz 1,
76344 Eggenstein-Leopoldshafen, Germany}
\author{Jan\,Jaroszynski}
%\email[Electronical address:\,]{jaroszy@magnet.fsu.edu}
\affiliation{Applied Superconductivity Center, National High Magnetic Field Laboratory, Florida State University, Tallahassee FL 32310, USA}
\author{Hidenori\,Hiramatsu}
%\email[Correspondence to:\,]{h-hirama@lucid.msl.titech.ac.jp}
\affiliation{Laboratory for Materials and Structures, Institute of Innovative Research, Tokyo Institute of Technology, Mailbox R3-1, 4259 Nagatsuta-cho, Midori-ku, Yokohama 226-8503, Japan}
\affiliation{Materials Research Center for Element Strategy, Tokyo Institute of Technology, Mailbox SE-6, 4259 Nagatsuta-cho, Midori-ku, Yokohama 226-8503, Japan}
\author{Bernhard\,Holzapfel}
\affiliation{Karlsruhe Institute of Technology, Institute for Technical Physics, Hermann-von-Helmholtz-Platz 1,
76344 Eggenstein-Leopoldshafen, Germany}
\author{Hideo\,Hosono}
%\email[Electronical address:\,]{hosono@msl.titech.ac.jp}
\affiliation{Laboratory for Materials and Structures, Institute of Innovative Research, Tokyo Institute of Technology, Mailbox R3-1, 4259 Nagatsuta-cho, Midori-ku, Yokohama 226-8503, Japan}
\affiliation{Materials Research Center for Element Strategy, Tokyo Institute of Technology, Mailbox SE-6, 4259 Nagatsuta-cho, Midori-ku, Yokohama 226-8503, Japan}
\date{\today}

\begin{abstract}
\noindent{\bf
High temperature (high-$T_{\rm c}$) superconductors like cuprates have superior critical current properties in magnetic fields over other superconductors. However, superconducting wires for high-field-magnet applications are still dominated by low-$T_{\rm c}$ Nb$_3$Sn due probably to cost and processing issues. The recent discovery of a second class of high-$T_{\rm c}$ materials, Fe-based superconductors, may provide another option for high-field-magnet wires. In particular, $AE$Fe$_2$As$_2$  ($AE$: Alkali earth elements, $AE$-122) is one of the best candidates for high-field-magnet applications because of its high upper critical field, $H_{\rm c2}$, moderate $H_{\rm c2}$ anisotropy, and intermediate $T_{\rm c}$. Here we report on in-field transport properties of P-doped BaFe$_2$As$_2$ (Ba-122) thin films grown on technical substrates (i.e., biaxially textured oxides templates on metal tapes) by pulsed laser deposition. The P-doped Ba-122 coated conductor sample exceeds a transport $J_{\rm c}$ of $10^5$\,A/cm$^2$ at 15\,T for both major crystallographic directions of the applied magnetic field, which is favourable for practical applications. Our P-doped Ba-122 coated conductors show a superior in-field $J_{\rm c}$ over MgB$_2$ and NbTi, and a comparable level to Nb$_3$Sn above 20\,T. By analysing the $E-J$ curves for determining $J_{\rm c}$, a non-Ohmic linear differential signature is observed at low field due to flux flow along the grain boundaries. However, grain boundaries work as flux pinning centres as demonstrated by the pinning force analysis.}
\end{abstract}

\maketitle
\subsection*{Introduction}
The discovery of Fe-based superconductors (FBS) by Kamihara $et$ $al.,$\,\cite{Kamihara} brought a huge impact to the physics community, since the compound consists of ferromagnetic Fe, which had been believed to be inevitably detrimental to the formation of Cooper pairs. To date, fundamental questions, such as mechanism of Cooper pairing and order parameter symmetry, are still under debate\,\cite{Peter}. On the other hand, this material class is attractive for applications. For instance, $AE$Fe$_2$As$_2$ ($AE$: Alkali earth elements, $AE$-122) and Fe(Se,Te) possess high upper critical fields ($H_{\rm c2}$) exceeding 50\,T and a low $H_{\rm c2}$ anisotropy close to 1 at low temperature\,\cite{Yamamoto,Lei}, which is favourable for high-field-magnet applications. Furthermore, Ba-122 shows less deterioration of critical current across grain boundaries (GBs)\,\cite{Katase-GB, Sakagami-GB} than YBa$_2$Cu$_3$O$_{7-\delta}$ (YBCO) and Bi-based cuprates. For Co-doped Ba-122, the critical GB misorientation angle ($\theta_{\rm c}$), where $J_{\rm c}$ starts to fall off exponentially, has been reported to be $6^\circ - 9^\circ$\,\cite{Lee,Katase-GB}. Even high angle GBs do not impede the current flow very much in sintered K-doped Ba-122 wires and bulks, if clean and well-connected GBs are realised\,\cite{Weiss-1, Weiss-2}. Additionally, Co-doped Ba-122 exhibits a high tolerance for large densities of flux pinning centres in the superconducting matrix, which leads to significant increase in critical current density ($J_{\rm c}$) and irreversibility field ($H_{\rm irr}$)\,\cite{Chiara-1}.

Another advantage of Ba-122, in particular P-doped Ba-122, is its inherently high $J_{\rm c}$. Putzke $et$ $al$. have reported on the enhancement of the vortex core energy of the flux lines at the quantum critical point (QCP) of the antiferromagnetic phase\,\cite{Putzke}. Indeed, even microstructurally clean and optimally P-doped Ba-122 epitaxial thin films, which were prepared by molecular beam epitaxy (MBE), exhibit a high self-field $J_{\rm c}$ of over 6\,MA/cm$^2$ at 4.2\,K\,\cite{Fritz-1}. Although excess magnetic Fe has been found to be harmful to superconductivity in Fe(Se,Te)\,\cite{Bendele}, Fe-rich P-doped Ba-122 thin films showed a higher self-field $J_{\rm c}$ of over 10\,MA/cm$^2$ at 4.2\,K, which is the highest value ever reported for FBS\,\cite{Sakagami-GB}. Whereas in the former case Fe is incorporated interstitially\,\cite{Sun}, in the latter case the Fe may form Fe-containing particles or regions with differing P-content, both acting as pinning centres\,\cite{Sakagami-GB}. Furthermore, the high $J_{\rm c}$ and low anisotropy P-doped Ba-122 thin films can be fabricated by tuning the processing conditions only, without any modification of the target material used in pulsed laser deposition (PLD)\,\cite{Sato-1}.

The aforementioned advantages of P-doped Ba-122 are very suitable for high-field-magnet applications. Indeed, P-doped Ba-122 thin films on technical substrates have been demonstrated as FBS coated conductors\,\cite{Hosono-IBAD, Sato-2}. To date, two kinds of technical substrates have been employed for FBS coated conductors: The cube-textured metal tapes with buffer layers (i.e., RABiTS)\,\cite{Norton} and the Hastelloy tape on which biaxially textured buffer layers are prepared by ion-beam-assisted-deposition (IBAD)\,\cite{Iijima}.

In contrast to Fe(Se,Te) coated conductors\,\cite{Si-IBAD,Si-Rabits}, transport properties of P-doped Ba-122 coated conductors in the presence of extremely high magnetic fields have not yet been reported. Here, we report on in-field transport properties of a P-doped Ba-122 thin film grown by PLD on metal substrate with biaxially textured MgO template (IBAD-MgO) in a wide range of temperature and DC magnetic field up to 35\,T. We employ IBAD-MgO template with a relatively large in-plane full width at half maximum (FWHM) value ($\Delta \phi_{\rm MgO}=8^\circ$), since it has been demonstrated by x-ray diffraction (XRD) and transmission electron microscopy (TEM) that the texture of MgO is transferred to the overlying P-doped Ba-122 film, generating dislocation networks\,\cite{Sato-2}. Such dislocation networks enhance the vortex pinning in P-doped Ba-122\,\cite{Sato-2}, since $\theta_{\rm c}$ is less than $9^\circ$\,\cite{Katase-GB}. Indeed, in-field $J_{\rm c}$ properties of our P-doped Ba-122 on IBAD-MgO with $\Delta \phi_{\rm MgO}=8^\circ$ were superior to those of the film on a template with $\Delta \phi_{\rm MgO}=4^\circ$\,\cite{Sato-2}. A high density of threading dislocations is very effective for improving $J_{\rm c}$ for $H\parallel c$ in a wide range of temperature and magnetic field even close to $H_{\rm irr}$. Despite the relatively large $\theta_{\rm c}$ of $6^\circ - 9^\circ$ for Ba-122,
$J_{\rm c}$ of our P-doped Ba-122 coated conductor with sharp FWHM values of both in-plane, $\Delta \phi_{\rm Ba-122}=5.7^\circ$, and out-of-plane misorientaion, $\Delta \omega_{\rm Ba-122}=1.2^\circ$ (see Supplemental Fig.\,S1) is limited by the GBs in the low field regime. However, at high field, it exceeds a transport $J_{\rm c}$ of $10^5$\,A/cm$^2$ at 15\,T for field applied in both main crystallographic directions. Our P-doped Ba-122 coated conductor sample shows superior in-field $J_{\rm c}$ properties over MgB$_2$ and NbTi, and a comparable level to Nb$_3$Sn above 20\,T.

\subsection*{Results}
\subsubsection{Resistivity measurements}
\begin{figure}
	\centering
		\includegraphics[width=10cm]{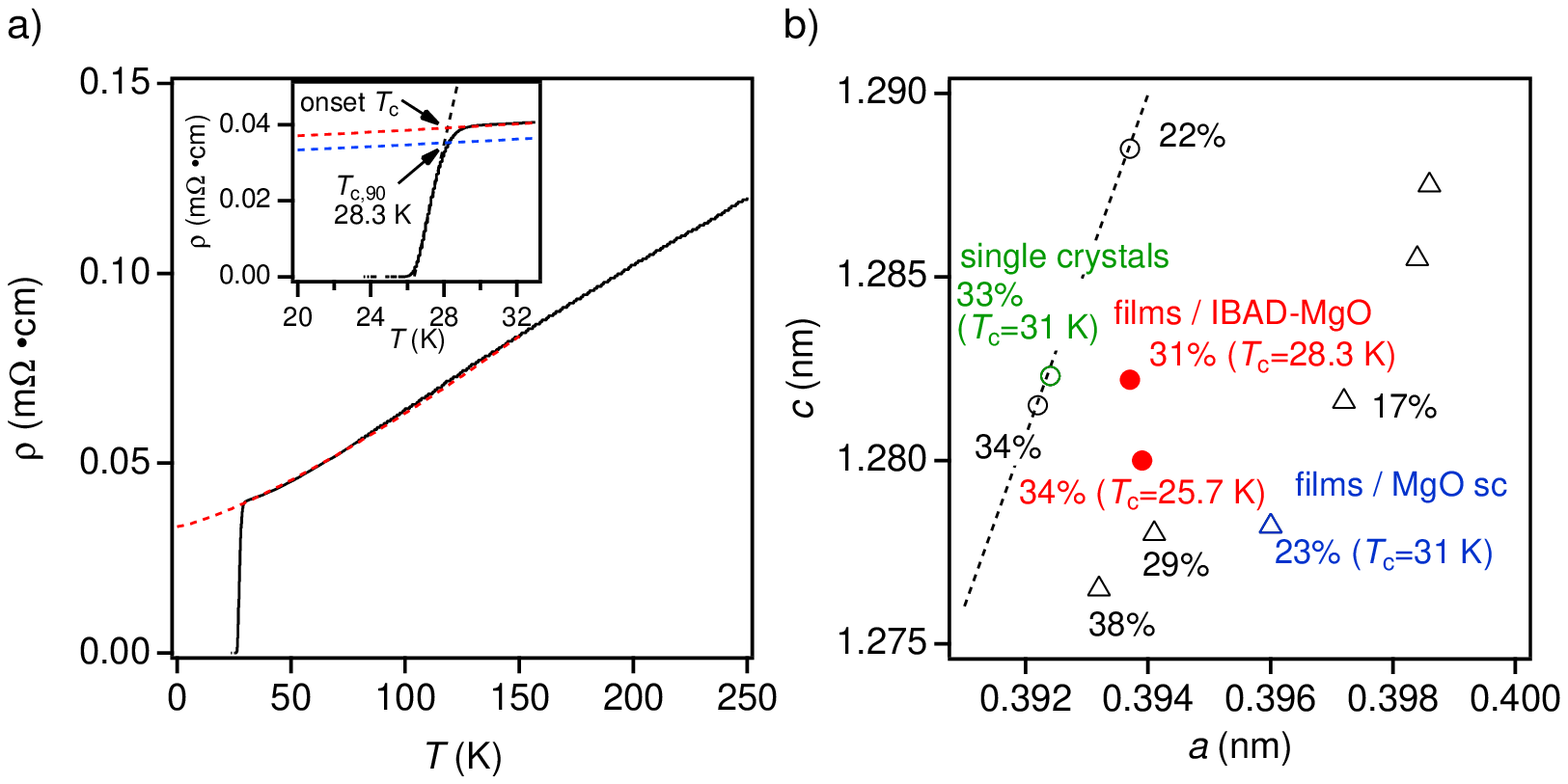}
		\vspace{4cm}
		\caption{{\bf Resistivity and the relationship between structural parameters and $T_{\rm c}$:} a) Temperature dependence of the resistivity measured in the absence of magnetic field. The dotted red line is a fit of the normal state resistivity curve in the range of $30<T<150$\,K using $\rho_{\rm n}=\rho_0+AT^n$. Inset: resistivity near the superconducting transition. The dotted blue line is a 90\% reduction of the fitting curve (red dotted line). A $T_{\rm c,90}$ of 28.3\,K was recorded. b) The relationship between lattice parameters $a$ and $c$ for P-doped Ba-122 single crystals\,\cite{Kasahara01} and thin films on MgO single crystal substrates\,\cite{Kawaguchi01} for various P contents. The highest-$T_{\rm c}$ values for single crystal and thin film are obtained by a P content of 33\% and 23\%, respectively. Lattice parameters $a$ and $c$ of our P-doped Ba-122 thin films on IBAD-MgO (two samples) are located between single crystals and thin films deposited on MgO single crystalline substrates. Both films have almost comparable $\Delta \phi_{\rm Ba122}$ and $\Delta \omega_{\rm Ba122}$ values.}
\label{fig:figure1}
\end{figure}

The normal-state resistivity $\rho_{\rm n}$ (Fig.\,\ref{fig:figure1}a) can be approximated by $\rho_{\rm n}=\rho_0+AT^n$ with an exponent $n$-value of 1.28, $\rho_0=3.32\times10^{-2}$\,m$\Omega \cdot$cm and $A=8.22\times10^{-5}$\,m$\Omega \cdot$cm/K$^{1.28}$ in the range of $30<T<150$\,K in accord with Ref.\,\onlinecite{Kasahara01}. Shibauchi $et$ $al$. have reported that the exponent $n$ is unity at the quantum critical point (QCP) of the antiferromagnetic phase, where the maximum $T_{\rm c}$ is observed at 33\% of P content for bulk single crystals\,\cite{Shibauchi}. Based on those results, we infer that the P content of our Ba-122 thin film on IBAD-MgO is different from the optimal level. Chemical analysis by electron probe microanalysis revealed a P content of 0.31, high enough to induce superconductivity with an onset $T_{\rm c}$ of 30\,K for Ba-122 single crystal\,\cite{Kasahara01}. The lower $T_{\rm c}$ (28.3\,K) of the P-doped Ba-122 coated conductor may be a consequence of epitaxial strain, since MgO single crystalline substrates induce in-plane tensile strain to Ba-122 films due to the lattice mismatch\,\cite{Kawaguchi01,Iida-strain}. The lattice parameters $a$ and $c$ of our P-doped Ba-122 coated conductors are located between the single crystals and thin films deposited on MgO single crystalline substrates (Fig.\,\ref{fig:figure1}b). The crystalline quality of IBAD-MgO affects mainly $\Delta \phi_{\rm Ba122}$ rather than $\Delta \omega_{\rm Ba122}$\,\cite{Sato-2}, changing the amount of the in-plane strain and hence $T_{\rm c}$.

\begin{figure}
	\centering
		\includegraphics[width=10cm]{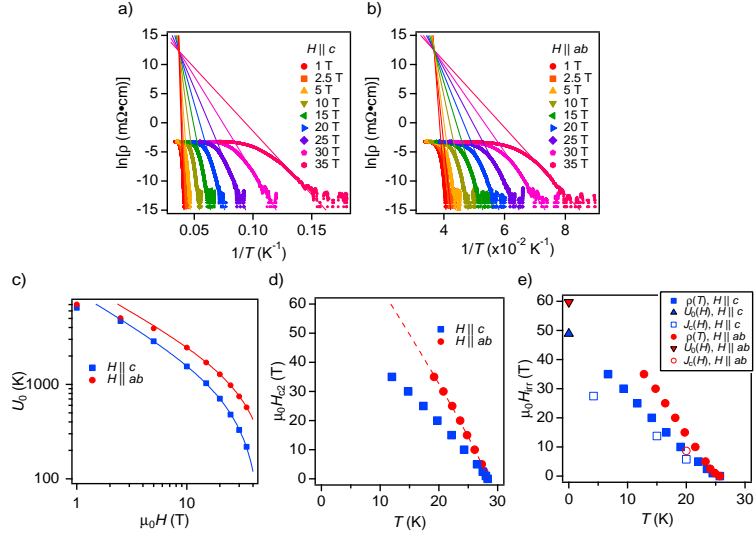}
		\vspace{4cm}
		\caption{{\bf Analysis of the activation energy for pinning potential, the upper critical field and the irreversibility field:} Arrhenius plots of the resistivity curves for a) $H\parallel c$ and b) $H\parallel ab$. c) Field dependence of the pinning potential for both main crystallographic directions. d) Temperature dependence of the upper critical field for both major directions. e) Temperature dependence of the irreversibility field for both major directions evaluated from $\rho(T,H)$ and $J_{\rm c}(T,H)$ measurements. Zero temperature $H_{\rm irr}$ for $H\parallel c$ and $\parallel ab$ evaluated from $U_0(H)$ is also shown.}
\label{fig:figure2}
\end{figure}

The linearity of the Arrhenius plots of $\rho(T,H)$ for both major crystallographic directions at a certain magnetic field (Figs.\,\ref{fig:figure2}a and \ref{fig:figure2}b) reveals thermally activated flux motion under the assumption of a linear $T$-dependence of the activation energy, $U_0(H)$\,\cite{Palstra} (See the method section).  It can be seen from Fig.\,\ref{fig:figure2}c that $U_0(H)$ for both $H\parallel c$ and $\parallel ab$ are well described by $H^{\alpha}(1-H/H^*)^\beta$ above 10\,T, which has been used for analysing polycrystalline MgB$_2$ samples by Thompson $et$ $al$\,\cite{Thompson}. $H^*$ is a characteristic field representing the irreversibility field at 0\,K\,\cite{Thompson,Jens-1}. The evaluated values for $H\parallel c$ and $\parallel ab$ are 48.9\,T and 59.7\,T, respectively (for $H\parallel c$ and $\parallel ab$ $\alpha=0.68$ and 0.64, and $\beta=1.1$ and 0.94). 

A linear fit for ${\rm ln}\rho{(H)}$ versus $U_0(H)$ using ${\rm ln}\rho_0(H)={\rm ln}\rho_{\rm 0f}+U_0(H)/T_{\rm c}$, where $\rho_{\rm 0f}$ is the prefactor, yields $T_{\rm c}$  of 26.9\,K for $H\parallel c$ and 27.2\,K for $H\parallel ab$, respectively (see Supplemental Fig.\,S2a). The $T_{\rm c}$ values evaluated by this method are slightly lower than the $T_{\rm c,90}$ (see Fig.\,\ref{fig:figure1}a). A plausible explanation for this difference is the increased transition width $\Delta T_{\rm c}$ due to the reduced texture quality compared to films on single crystal substrates or single crystal samples.

$H_{\rm c2}(T)$ was evaluated from the linear presentations of Figs.\,\ref{fig:figure2}a and \ref{fig:figure2}b (see Supplementary Fig.\,S2b and S2c) applying a $\rho_{\rm n,0.9}=0.9\rho_{\rm n}$ resistivity criterion, where $\rho_{\rm n,0.9}$ is the normal state resistivity $\rho_{\rm n}$ at 28.5\,K. Shown in Fig.\,\ref{fig:figure2}d is $H_{\rm c2}$ for $H\parallel c$ and $\parallel ab$. The dotted line in Fig.\,\ref{fig:figure2}d is the fitting curve using $(1-T/T_{\rm c})^k$. An exponent $k$ of 0.9 was obtained for $H\parallel ab$, which is far from the expected value of 0.5 for layered compounds limited by Pauli pair breaking at given $T$ close to the dimensional crossover temperature\,\cite{Uher,Klemm,Chiara-2}, which confirms that P-doped Ba-122 is a 3D superconductor. Because of the lack of low temperature data, it is not possible to fit the $H_{\rm c2}(T)$ (and $H_{\rm c2}(\theta)$, shown later unambiguously) with a proper model for FBS\,\cite{Gurevich-1,Gurevich-2}.

The temperature dependence of the irreversibility field, $H_{\rm irr}(T)$ (Fig.\,\ref{fig:figure2}e) was evaluated from $\rho(T,H)$ measurements using a resistivity criterion of $\rho_{\rm c}=E_{\rm c}/J_{\rm c,100}=1.0\time10^{-8}\,{\rm \Omega \cdot cm}$, where $E_{\rm c}$ is the electric field criterion ($1\,{\rm \mu V/cm}$) for determining $J_{\rm c}$ from $E-J$ measurements and $J_{\rm c,100}$ is the criterion ($100\,{\rm A/cm^2}$) for determining $H_{\rm irr}$ from $J_{\rm c}(H)$ measurements (see Supplementary Fig.\,S2d and S2e). The $H_{\rm irr}$ data at\,0 K are estimated from the Arrhenius plots and they appear to match the low temperature limit of the $H_{\rm irr}$ data directly determined from the $\rho(T,H)$ using the $\rho_{\rm c}$ criterion. For comparison, $H_{\rm irr}(T)$ determined from $J_{\rm c}(H)$ is also plotted in Fig.\,\ref{fig:figure2}e showing some differences with the values estimated from $\rho(T,H)$. A plausible reason is a different frequency of the applied current used in those investigations\,\cite{Pan}.

\begin{figure}
	\centering
		\includegraphics[width=10cm]{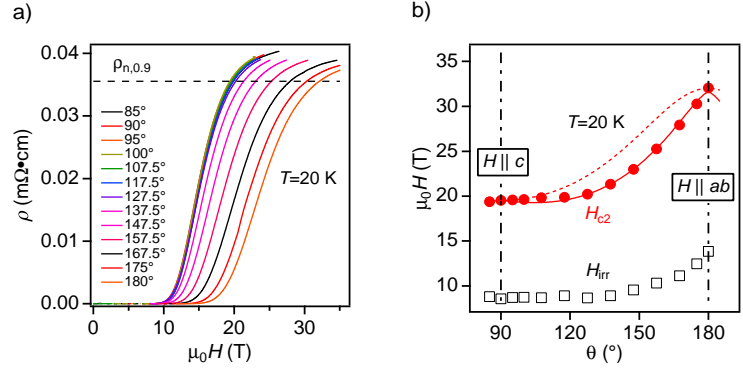}
		\vspace{4cm}
		\caption{{\bf Angular dependence of the upper critical field and the irreversibility field:} a) Angular dependence of magnetoresistivity [$\rho(H)$] at 20\,K up to 35\,T. b) Anglar dependence of the upper critical field and the irreversibility field at 20\,K. The solid line is the fitting curve using eq. (1) with $\delta=1.47$ and $\gamma=1.62$. The dotted line is the AGL dependence.}
\label{fig:figure3}
\end{figure}

The angular dependence of $H_{\rm c2}$ at 20\,K, which was derived from $\rho(H)$ curves at constant angles with $\rho_{\rm n,0.9}$ (Fig.\,\ref{fig:figure3}a) shows a minimum at $\theta=90^\circ$ ($H \parallel c$) and a maximum at $\theta=180^\circ$ ($H \parallel ab$), as shown in Fig.\,\ref{fig:figure3}b. The single-band anisotropic Ginzburg-Landau (AGL) theory\,\cite{AGL}, $H_{\rm c2}(\theta)=H_{\rm c2}(90^\circ)(\rm sin^2(\theta)+cos^2(\theta)/\gamma^2)^{-0.5}$ with $\gamma=H_{\rm c2}(180^\circ)/H_{\rm c2}(90^\circ)$ (dotted line in Fig.\,\ref{fig:figure3}b), cannot describe the measured $H_{\rm c2}(\theta)$ due to the multi-band nature of this material, similarly to Co-doped Ba-122\,\cite{Jens-1}. A fairly good description of the data is, however, achieved by the empirical formulae\,\cite{Jens-1}, 

\begin{eqnarray}
H_{\rm c2}(\theta)=H_{\rm c2}(90^\circ)\times \epsilon(\theta, \gamma, \delta),\,\,\, \epsilon(\theta, \gamma, \delta)=\left(\left| {\rm sin}\theta \right|^\delta+\left|\frac{{\rm cos}\theta}{\gamma}\right|^\delta\right)^{-\frac{1}{\delta}}
\end{eqnarray}

\noindent with $\delta=1.47$ and $\gamma=1.62$ (solid line). The parameter $\gamma$ is the $H_{\rm c2}$ anisotropy, whereas $\delta$ is a measure for the $ab$-peak width whose physical meaning is still unclear. These two values will be used later for scaling the angular dependence of $J_{\rm c}(\theta)$ data.

The angular dependence of $H_{\rm irr}$ at 20\,K derived using the same resistivity criterion $\rho_{\rm c}=1.0\time10^{-8}\,{\rm \Omega \cdot cm}$ shows almost the same trend as $H_{\rm c2}(\theta)$. Unlike the angular dependence of $J_{\rm c}$ (see next section), no clear peak at $\theta=90^\circ$ ($H \parallel c$) is observed in $H_{\rm irr}(\theta)$.

\subsubsection{In-field critical current density $J_{\rm c}(T, H, \theta)$}
\begin{figure}
	\centering
		\includegraphics[width=10cm]{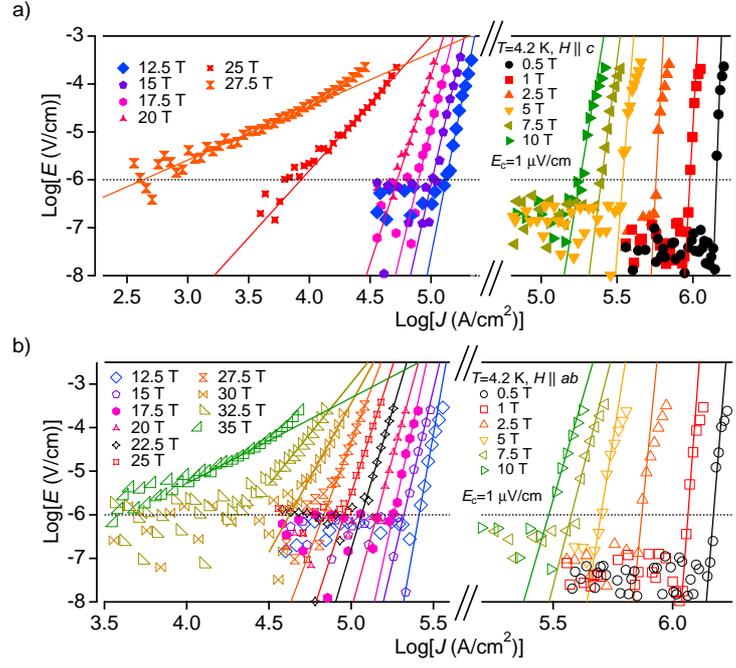}
		\vspace{4cm}
		\caption{{\bf $E-J$ characteristics:} $E-J$ curves for P-doped Ba-122 measured at 4.2\,K for a) $H\parallel c$ in the range of $0.5<\mu_0H<27.5$\,T and b) for $H\parallel ab$ in the range of $0.5<\mu_0H<35$\,T. The electric field criterion of 1\,$\mu {\rm V/cm}$ for evaluating $J_{\rm c}$ is also shown. A NOLD signature can be identified at low field by the deviation from the linear trend observed at high $E$.}
\label{fig:figure4}
\end{figure}

The $E-J$ curves of the P-doped Ba-122 coated conductor sample at 4.2\,K (Fig.\,\ref{fig:figure4}) show different behaviour at high and low magnetic fields for both major field directions. Up to 10\,T they exhibit a non-Ohmic linear differential (NOLD) signature (i.e., $E$ is linearly changing with $J$ in linear scale, see Supplemental Fig.\,S3), indicative of $J_{\rm c}$ limitation by GBs\,\cite{Verebelyi}. Here NOLD behaviour is due to viscous flux flow along the GBs\,\cite{Diaz-1}. On the other hand, NOLD signature is almost absent above 12.5\,T, suggesting that $J_{\rm c}$ is limited by intra-grain depinning of flux lines. This pinning crossover field is observed to decrease with increasing temperature (not shown), which is consistent with the cuprate YBCO reported in Ref.\,\onlinecite{Daniels,Fernandez}.

\begin{figure}
	\centering
		\includegraphics[width=10cm]{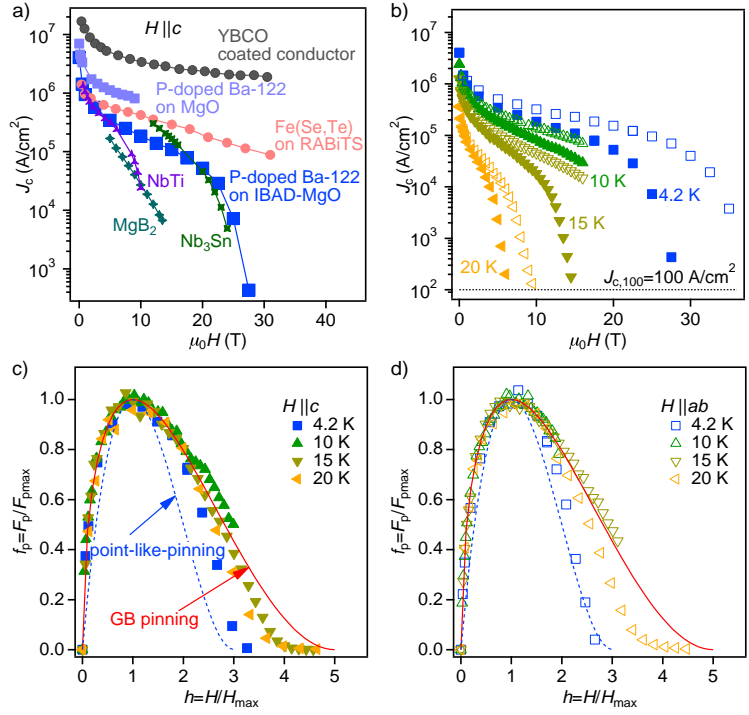}
		\vspace{4cm}
		\caption{{\bf Field dependence of $J_{\rm c}$ and analysis of the flux pinning density:} a) $J_{\rm c}-H$ properties of the P-doped Ba-122 coated conductor sample at 4.2\,K for $H\parallel c$ in comparison to the data for P-doped Ba-122 on MgO single crystalline substrate\,\cite{Sato-1}, Fe(Se,Te) on RABiTS\,\cite{Si-Rabits}, YBCO coated conductor\,\cite{Xu}, MgB$_2$\,\cite{GZLi}, NbTi\,\cite{Boutboul,Kanithi}, and Nb$_3$Sn\,\cite{Parrell-1,Parrell-2}. b) $J_{\rm c}-H$ properties of the P-doped Ba-122 coated conductor sample at various temperatures for both $H\parallel c$ (closed symbols) and $H\parallel ab$ (open symbols).  $J_{\rm c,100}$ is the criterion ($100\,{\rm A/cm^2}$) used for determining $H_{\rm irr}$. c) and d) The normalised pinning force $f_{\rm p}$ as a function of reduced field $h$. The solid line is $f_{\rm p}=\frac{25}{16}h^{0.5}(1-\frac{h}{5})^2$ for GB pinning, and the blue dotted line represent point-like pinning, $f_{\rm p}=\frac{9}{4}h(1-\frac{h}{3})^2$.}
\label{fig:figure5}
\end{figure}

Figure\,\ref{fig:figure5}a compares $J_{\rm c}(H)$ for P-doped Ba-122 on IBAD-MgO for $H\parallel c$ at 4.2\,K with P-doped Ba-122 on MgO single crystalline substrate\,\cite{Sato-1}, Fe(Se,Te) on RABiTS\,\cite{Si-Rabits}, YBCO coated conductor\,\cite{Xu}, MgB$_2$\,\cite{GZLi}, NbTi\,\cite{Boutboul,Kanithi}, and Nb$_3$Sn\,\cite{Parrell-1,Parrell-2}. Pinning-improved YBCO 2nd-generation (2G) tape shows the highest $J_{\rm c}$ at entire magnetic fields; however, a well textured template is necessary. The P-doped Ba-122 coated conductor exceeds a self-field $J_{\rm c}$ of 4\,MA/cm$^2$ and maintains a high $J_{\rm c}$ value of 50\,kA/cm$^2$ at 20\,T. For the entire field range, $J_{\rm c}$ of P-doped Ba-122 coated conductor sample is larger than for MgB$_2$ and NbTi. Above 20\,T, the P-doped Ba-122 coated conductor sample shows comparable properties to Nb$_3$Sn. Although lower-field $J_{\rm c}$ of P-doped Ba-122 on IBAD-MgO is higher than that of Fe(Se,Te) on RABiTS, the latter shows the better performance at medium and high fields. Figure\,\ref{fig:figure5}b summarises $J_{\rm c}(H)$ for P-doped Ba-122 on IBAD-MgO for both crystallographic directions at various temperatures. At intermediate fields $J_{\rm c}$ for the two directions is comparable, indicative of the presence of correlated pinning along the $c$-axis.

By analysing the pinning force density $F_{\rm p}=\mu_0H\times J_{\rm c}$, information on vortex pinning can be obtained. In general, the normalised pinning force, $f_{\rm p}=F_{\rm p}/F_{\rm p,max}$, is plotted as a function of reduced field $h_1=H/H_{\rm irr}$ at a given temperature for high-$T_{\rm c}$ superconductors. However, we plot $f_{\rm p}$ as a function of $h=H/H_{\rm max}$, where $H_{\rm max}$ is the field at which $F_{\rm p}$ shows the maximum\,\cite{Civale,Qin,Higuchi,Paturi}, since $J_{\rm c}$ could not be measured up to $H_{\rm irr}$ at all temperatures. As can be seen in Fig.\,\ref{fig:figure5}c, the $f_{\rm p}$ curves at different temperatures for $H\parallel c$ almost fall onto a master curve in the range of $0<h<3$ described by
  
\begin{eqnarray}
f_{\rm p}=\frac{25}{16}h^{0.5}(1-\frac{h}{5})^2
\end{eqnarray}  

\noindent This formula is analogous to $h_1^p(1-h_1)^q$ ($p=0.5$ and $q=2$) found by Dew-Hughes\,\cite{Hughes} for pinning by planar defects such as GB and twin boundaries, and by Kramer for line defect arrays\,\cite{Kramer}. In high-$T_{\rm c}$ superconductors with extremely short coherence lengths $\xi$, a further classification of the defect size with respect to $\xi$ is necessary. It has been recently found by Paturi $et$ $al.$ that the exponent $p$ is 0.5 irrespective of $q$ for a defect size of the order of $\xi$ and especially for dislocations in undoped YBCO films\,\cite{Paturi}. On the contrary, $p$ increases towards 1 with increasing defect size. This confirms the finding that pinning in our sample is dominated by the dislocations with nano-size. Here, it should be noted that a sign of NOLD signature does not contradict GB pinning. In fact it has been reported for YBCO that the dislocations in GBs can work as vortex pinning centres\,\cite{Diaz-2, Heinig}. The flux preferentially flows across the dislocation cores in the GB plane, which explains the $E-J$ curves with NOLD sign.

Abrikosov-Josephson vortices (AJV) are present in low-angle GBs in both YBCO\,\cite{Gurevich-3} and FBS. Unlike Josephson vortices (JV), AJV have normal cores and can be trapped by flux pinning. Furthermore, the presence of an interaction between Abrikosov vortices (AV) in the grain and AJV at the GBs has been experimentally found in Ref.\,\onlinecite{Palau}: an increase in pinning potential for AV leads to the enhancement of the pinning potential for AJV.

For $H\parallel ab$ the $f_{\rm p}$ curves at both 10 and 15\,K follow well the GB pinning line (red solid line) up to 16\,T (corresponding to $h=2$ and 3.2 in Fig.\,\ref{fig:figure5}d, respectively). In contrast, $f_{\rm p}$ at 20\,K neither follows the GB pinning  nor point-like pinning (red solid and blue dashed lines, respectively) in high field regime, although the $f_{\rm p}$ curve lies on the GB pinning line below $h<2$. Similarly, the $f_{\rm p}$ curve at 4.2\,K follows the GB pinning curve up to $h<1.5$ and then approaches the point-like pinning curve beyond $h>1.5$. Hence, differently from the $H\parallel c$ case, the dominant pinning mechanism for $H\parallel ab$ is varying with temperature and field strength.

\begin{figure}
	\centering
		\includegraphics[width=10cm]{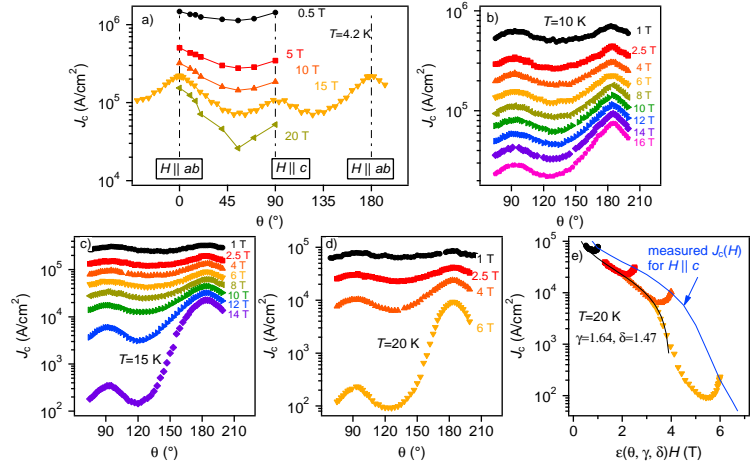}
		\vspace{4cm}
		\caption{{\bf Field and orientation dependence of $J_{\rm c}$:} Angular dependence of $J_{\rm c}$ measured at a) 4.2\,K, b) 10\,K, c) 15\,K, and d) 20\,K. e) Scaling behaviour of $J_{\rm c}(\theta)$ at 20\,K as a function of effective field, $\epsilon(\theta, \gamma, \delta)\times \mu_0H$. $\delta=1.47$ and $\gamma=1.62$ evaluated from the $H_{\rm c2}(\theta)$ were used. Blue solid line is the measured $J_{\rm c}-H$ for $H\parallel c$ at 20\,K.}
\label{fig:figure6}
\end{figure}

The angular dependence of the critical current density, $J_{\rm c}(\theta)$ (Fig\ref{fig:figure6}a-d), shows two distinct peaks: a relatively sharp peak at $H\parallel ab$ and a broad maximum at $H\parallel c$, which arises from the network of threading dislocations comprising the low-angle GBs\,\cite{Sato-2}. Surprisingly, the $c$-axis peaks [$J_{\rm c}(90^\circ)$] remain visible even close to $H_{\rm irr}$ at all temperatures. Unlike single band superconductors, the anisotropy of coherence length, $\gamma_\xi=\xi_{ab}/\xi_{c}$, and penetration depth, $\gamma_\lambda=\lambda_{c}/\lambda_{ab}$, of FBS exhibit opposite behaviour with temperature\,\cite{Konczykowski}. For an optimally doped Ba-122 system, $\gamma_\lambda>\gamma_\xi$ holds at all temperature. In this case even occasional uncorrelated  defects slightly larger than $\xi$ yield a strong $c$-axis pinning\,\cite{Beek}. Such an effect in combination with threading dislocations along the $c$-axis may enhance enormously the average pinning potential for applied fields parallel to the $c$-axis.

Shown in Fig.\,\ref{fig:figure6}e is the scaling behaviour of $J_{\rm c}(\theta)$ as a function of the effective field [i.e., $\epsilon(\theta, \gamma, \delta)\times \mu_0H$] at 20\,K. Here $\delta=1.47$ and $\gamma=1.62$ were used as obtained by the $H_{\rm c2}(\theta)$ fit. As can be seen, all $J_{\rm c}(\theta)$ curves collapse onto a master curve in a wide angular range around $H\parallel ab$. Differences between the master curve and the measured $J_{\rm c}(H)$ for $H\parallel c$ are correlated pinning contributions. Here we emphasise that the $J_{\rm c}$ peak at $\theta=180^\circ$ is fully determined by the electronic anisotropy at 20\,K and no intrinsic pinning or pinning by planar defects is observed.
 
\subsection*{Discussions and conclusions}
In order to realise FBS coated conductors, high $J_{\rm c}$ values with low anisotropy in high fields are necessary. $J_{\rm c}$ of our P-doped Ba-122 coated conductor nearly reached the practical level of $\sim$0.1\,MA/cm$^2$ at 15\,T for any applied field directions at 4.2\,K [see Fig.\,\ref{fig:figure5}a)], which shows superior properties over MgB$_2$ and NbTi. Above 20\,T the level of $J_{\rm c}$ is comparable to Nb$_3$Sn.  Additionally, the intrinsic anisotropy estimated at 20\,K from the $H_{\rm c2}$ data is below 2. Moreover, the correlated defects increase $J_{\rm c}$ for $H\parallel c$ substantially suppressing the effective $J_{\rm c}$ anisotropy.

As stated above, the inequality of $\xi$ and $\lambda$ anisotropy in combination with a large density of threading dislocations along the $c$-axis significantly enhances the average pinning potential. It is worth mentioning that the population of threading dislocations can be controlled by the processing conditions only, without any modification of the PLD target\,\cite{Sato-1}.

Compared to optimally P-doped Ba-122 films on MgO single crystal substrates by MBE\,\cite{Fritz-1} and PLD\,\cite{Sato-1}, the level of $J_{\rm c}$ of the P-doped Ba-122 coated conductor still needs to be improved. Film stoichiometry especially for P content should be controlled precisely. As stated before, the P content of our Ba-122 film slightly differs from the optimal level, where the QCP causes a sharp maximum for the vortex core energy\,\cite{Putzke}. As a consequence, the slight deviation from the optimal P level in our sample results in a lower vortex core energy, which directly reduces $J_{\rm c}$.

Unlike in electron and hole doped Ba-122 systems, aliovalent disorder that contributes to pinning in the Co or K cases is absent in P-doped Ba-122. However, $J_{\rm c}$ can be further enhanced by introducing growth defects (e.g. intragrain dislocations since the PLD processing
conditions strongly affect their density\,\cite{Sato-1}) and artificial structures (e.g. nanoparticles).
Moreover, the thermal conductivity of single crystalline MgO is different from that of IBAD-MgO template, which infers the optimum deposition temperature may change.

The introduction of artificial pinning centres is effective for further improvement of $J_{\rm c}$. In fact, Miura $et$ $al.$ have reported the introduction of BaZrO$_3$ into P-doped Ba-122 matrix\,\cite{Miura} in analogy to the addition of BaZrO$_3$ to YBCO. Hence, a combination of the introduction of artificial pinning centres and the precise control of P content will yield better performing P-doped Ba-122 coated conductors.

An attempt to fabricate a long length P-doped Ba-122 coated conductor has started quite recently. As a result, a 15\,cm long P-doped Ba-122 coated conductor has been realised by PLD using a reel-to-reel system\,\cite{Hosono-IBAD}. Albeit the resultant P-doped Ba-122 showed a small self-field $I_{\rm c}$ of 0.47\,mA (corresponding to a $J_{\rm c}$ of $4.7\times 10^4\,{\rm A/cm^2}$) at 4.2\,K, an improvement of $I_{\rm c}$ is foreseen by applying the aforementioned methods.

In summary, we have investigated in-field transport properties of a P-doped Ba-122 thin film grown by PLD on technical substrate in a wide range of temperature and DC magnetic field up to 35\,T. The P-doped Ba-122 coated conductor exceeds a transport $J_{\rm c}$ of $10^5$\,A/cm$^2$ at 15\,T for both major crystallographic directions of the applied field. Additionally, the $J_{\rm c}$ peaks for $H\parallel c$ remain visible even close to $H_{\rm irr}$ at all temperatures by the enhanced vortex pinning due to the combination of large population of threading dislocations and the inequality of $\xi$ and $\lambda$ anisotropy. This leads to a lower $J_{\rm c}$ anisotropy. By analysing pinning force densities, we established that the GB pinning contribution is dominant for $H\parallel c$, whereas for $H\parallel ab$, the dominant pinning is varying with temperature. The results obtained through this study are considered promising for future high-field-magnet applications of $AE$-122 systems.

\section*{Methods}
\subsection*{Growth of the P-doped Ba-122 film and structural characterisation}
The P-doped Ba-122 thin film of 185 \,nm thickness was grown by pulsed laser deposition on an IBAD-MgO Hastelloy metal-tape substrate supplied by iBeam Materials, Inc\,\cite{Sheehan}. The stacking structure of the IBAD-MgO substrate as shown in ref.\,\onlinecite{Sato-2} consists of first a planarising bottom bed-layer amorphous Y$_2$O$_3$ on the Hastelloy, second a biaxially textured MgO layer formed by IBAD, and a top homoepitaxial MgO layer. The IBAD-MgO substrate with a large in-plane distribution angle of $\Delta \phi_{\rm MgO}=8^\circ$ was investigated because higher $J_{\rm c}$ with isotropic properties can be achieved compared to the film on the well in-plane-aligned IBAD-MgO metal-tapes (i.e., $\Delta \phi_{\rm MgO}=4^\circ$)\,\cite{Sato-2}. A polycrystalline BaFe$_2$(As$_{0.65}$P$_{0.35}$)$_2$ disk was used as the PLD target. We employed a higher growth temperature of 1200\,$^\circ$C than for optimised P-doped Ba-122 films on MgO single-crystal substrates (1050\,$^\circ$C)\,\cite{Sato-1}, since the P concentration increases with increasing growth temperature for a given target composition. As expected, a higher P concentration closer to the optimum P concentration than in previous studies was achieved\,\cite{Sato-1,Sato-2}. The other growth parameters [e.g., the excitation source and the laser fluence of the second harmonics (wavelength: 532\,nm) of a Nd-doped yttrium-aluminum-garnet pulsed laser and 3\,J/cm$^2$, respectively] were the same as reported in Ref.\,\onlinecite{Sato-1}.

To determine the crystalline phases, $\omega$-coupled 2$\theta$ scan X-ray diffraction measurements were performed. The asymmetric 103 diffraction of the P-doped Ba-122 film was measured to confirm the in-plane crystallographic four-fold symmetry without in-plane rotational domains. The crystallinity of the film was characterised on the basis of the full widths at half maximum (FWHMs) of the out-of-plane 004 ($\Delta \omega$) and the in-plane 200 rocking curves ($\Delta \phi$). The results of those XRD measurements can be found in Supplementary Information Fig.\,S1. The chemical composition was determined with an electron-probe microanalyser. The acceleration voltage of the electron beam was optimised while monitoring the Ni K$\alpha$ spectrum to avoid the matrix effect from the Ni-containing Hastelloy metal-tapes.

\subsection*{In-plane transport measurements}
A small bridge of 15\,$\mu$m width and 500\,$\mu$m length was patterned by photolithography, followed by ion-beam etching. Au electrodes with 50\,nm thickness were formed by sputtering and lift-off. Transport properties using the resultant bridge were measured by a standard four-probe method.

The temperature dependence of the resistivity of the P-doped Ba-122 coated conductor shows a $T_{\rm c,90}$ of 28.3\,K (Fig.\,\ref{fig:figure1}a), which is about 3\,K lower than that of the optimally P-doped Ba-122 single crystals. $T_{\rm c,90}$ is defined as the intersection between the steepest slope of the superconducting transition and a 90\% reduction of the fit of the normal state resistivity using $\rho_{\rm n}=\rho_0+AT^n$. On the other hand, the onset $T_{\rm c}$ is defined as the intersection between the fit curve as stated above and the steepest slope of the superconducting transition. The difference between $T_{\rm c,90}$ and the onset $T_{\rm c}$ is negligible.

The activation energy $U_0(H)$ for vortex motion was evaluated by the temperature dependence of the resistivity measurements in various field strengths up to DC 35\,T at the National High Magnetic Field Laboratory, Tallahassee, FL, USA. According to the model of thermally activated flux flow\,\cite{Palstra}, the slope of linear fit yields the pinning potential for vortex motion at given fields (Fig.\,\ref{fig:figure2}c). On the assumption that $U(T,H)=U_0(H)(1-T/T_{\rm c})$, both equations, ${\rm ln}\rho(T,H)= {\rm ln}\rho_0(H)-U_0(H)/T$ and ${\rm ln}\rho_0(H)={\rm ln}\rho_{\rm 0f}+U_0(H)/T_{\rm c}$, are obtained, where  $\rho_{\rm 0f}$ is a prefactor.

In order to further understand the $H_{\rm c2}$ anisotropy for a P-doped Ba-122 coated conductor sample, the angular dependence of the magnetoresistivity was measured at 20\,K. Using the same constant criterion $\rho_{\rm n,0.9}$ for evaluating $H_{\rm c2}$, the angular dependent upper critical field [$H_{\rm c2}(\theta)$] was derived (Fig.\,\ref{fig:figure3}b).

A criterion of 1\,$\mu {\rm V/cm}$ was employed for evaluating $J_{\rm c}$. In $J_{\rm c}$ measurement, the magnetic field was always applied in the maximum Lorentz force configuration. Low-field measurements were performed in a Quantum Design physical property measurement system (PPMS) in magnetic fields up to 16\,T. For high field measurements up to DC 35\,T, the experiments were conducted at the National High Magnetic Field Laboratory, Tallahassee, FL, USA.

\subsection*{Acknowledgement} A portion of this work was performed at the National High Magnetic Field Laboratory, which was supported by National Science Foundation Cooperative Agreement No. DMR-1157490, and the State of Florida. The work at Tokyo Institute of Technology was supported by the Ministry of Education, Culture, Sports, Science and Technology (MEXT) through Element Strategy Initiative to Form Core Research Center. K.I. acknowledges support by the Japan Society for the Promotion of Science (JSPS) Grant-in-Aid for Scientific Research (B) Grant Number 16H04646. H.Hi was also supported by JSPS for Young Scientists (A) Grant Number 25709058, JSPS Grant-in-Aid for Scientific Research on Innovative Areas Nano Informatics (Grant Number 25106007), and Support for Tokyotech Advanced Research (STAR). We acknowledge support by Deutsche Forschungsgemeinschaft and Open Access Publishing Fund of Karlsruhe Institute of Technology.

\subsection*{Authors contribution} K.I., C.T., J.H., H.S. and H.Hi. designed the study and wrote the manuscript together with J.J. and H.Ho. Thin films preparation, structural characterisations and micro bridge fabrications were carried out by H.S. and H.Hi.\,~ K.I., C.T., J.H. and J.J. conducted high field transport measurements. C.T., H.S. and H.Hi. performed low field transport measurements. K.I., C.T., H.Hi, and H.Ho. supervised the projects. All authors discussed the results and implications and commented on the manuscript at all stages.

\section*{Additional information}
The authors declare no competing financial interests. Correspondence and requests for materials should be addressed to K.I.

\clearpage

\clearpage

\clearpage

\clearpage

\clearpage

\clearpage

\clearpage

\begin{center}
\textbf{\large Supplemental Information}
\end{center}

\noindent
\textbf{\large High-field transport properties of a P-doped BaFe$_2$As$_2$ film on technical substrate}

\setcounter{equation}{0}
\setcounter{figure}{0}
\setcounter{table}{0}
\setcounter{page}{1}

\makeatletter
\renewcommand{\theequation}{S\arabic{equation}}
\renewcommand{\thetable}{S\arabic{table}}
\renewcommand{\thefigure}{S\arabic{figure}}
\renewcommand{\bibnumfmt}[1]{[S#1]}
\renewcommand{\citenumfont}[1]{S#1}

\subsection{Structural characterisation}

\begin{figure}[h]
	\centering
		\includegraphics[width=9cm]{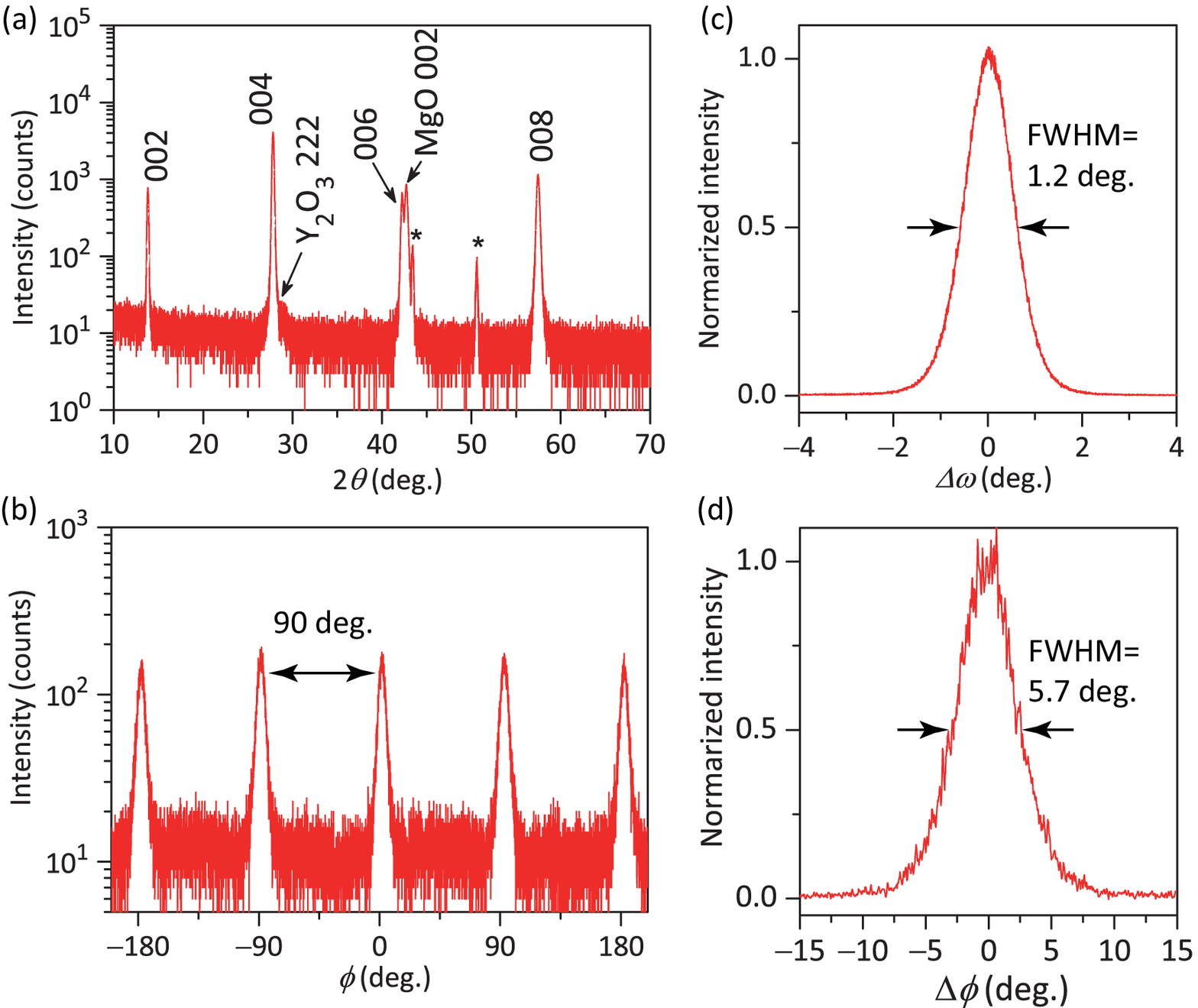}
		\caption{{\bf X-ray diffraction (XRD) patterns of a P-doped Ba-122 film on an IBAD-MgO substrate with in-plane misorientation angles of $\Delta \phi_{\rm MgO}=8^\circ$.} (a) $\omega$-coupled 2$\theta$ scan for out-of-plane reflections. The asterisks indicate the diffraction peaks from the IBAD-MgO substrate. (b) $\phi$ scan of the asymmetric 103 diffraction. Intensity-normalised rocking curves of (c) the out-of-plane 004 and (d) the in-plane 200 diffraction. These XRD data indicate that (i) The planarising amorphous Y$_2$O$_3$ bed-layers in the IBAD-MgO substrate slightly crystallised as observed at $2\theta=29^\circ$ due to high temperature growth at 1200\,$^\circ$C. (ii) The P-doped Ba-122 film heteroepitaxially grew on the IBAD-MgO substrate with the orientation relation of Ba-122[001]$\|$IBAD-MgO[001] out-of-plane and Ba-122[100]$\|$IBAD-MgO[100] in-plane without extra in-plane rotational domains. (iii) The FWHM value of out-of-plane diffraction (i.e., tilting of crystallites) is $\Delta \omega_{\rm Ba122}=1.2^\circ$, which is comparable to that of the previous study\,\cite{Sato-12}, whereas that ($\Delta \phi_{\rm Ba122}=5.7^\circ$) of in-plane diffraction (i.e., twisting of crystallites) is slightly improved from that ($\Delta \phi_{\rm Ba122}=8^\circ$) of the previous study mainly because of higher growth temperature employed in this study\,\cite{Sato-11,Sato-12}. The $\Delta \phi_{\rm Ba122}=5.7^\circ$ is smaller than that of IBAD-MgO ($\Delta \phi_{\rm MgO}=8^\circ$) due probably to self epitaxy effect during PLD growth of the Ba-122 film.}
\label{fig:figures1}
\end{figure}

\clearpage
\subsection{Resistivity curves for determining $H_{\rm c2}$ and $H_{\rm irr}$}
\begin{figure}[h]
	\centering
		\includegraphics[width=8cm]{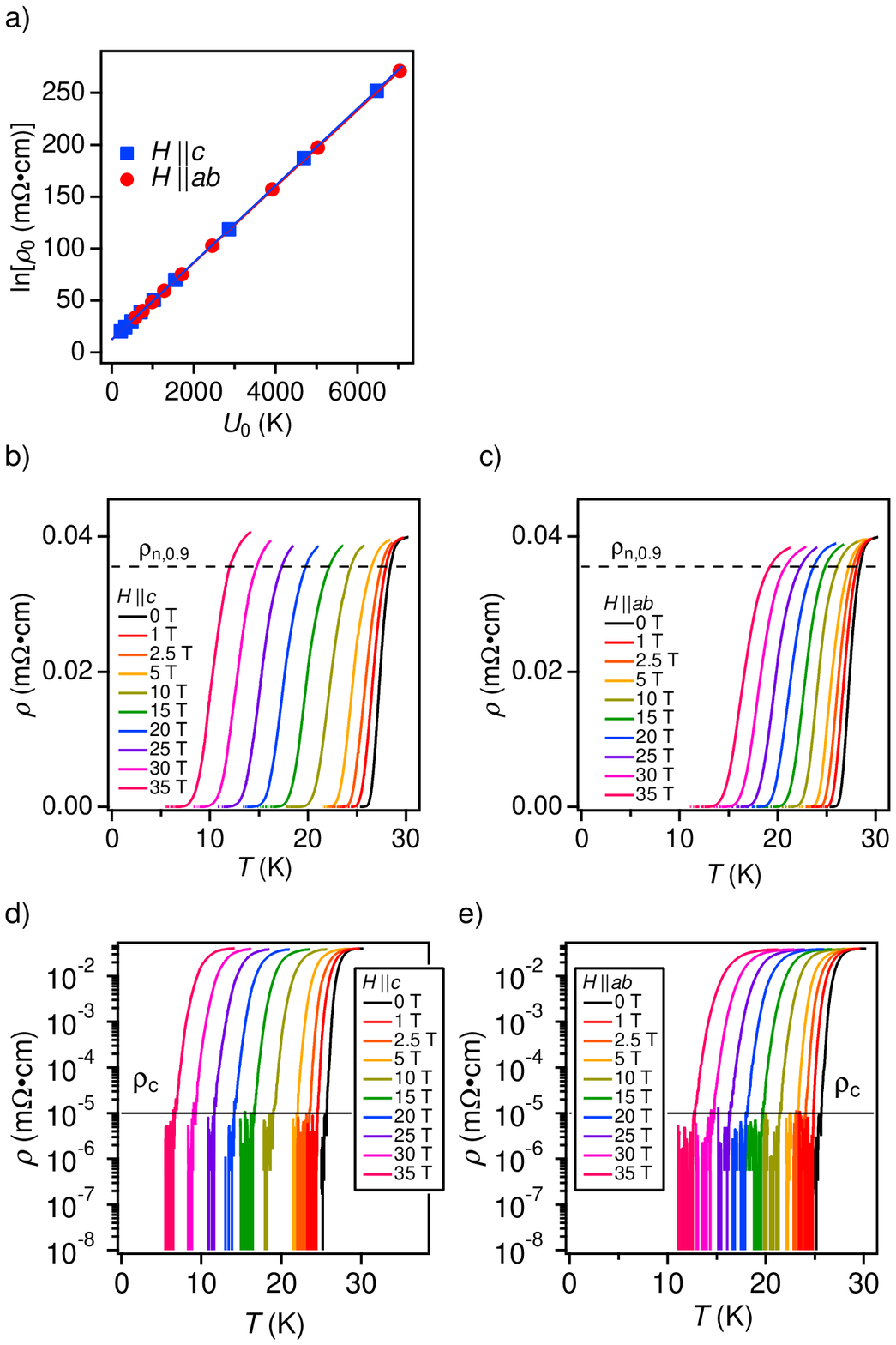}
		\vspace{1.5cm}
		\caption{{\bf Resistivity curves for determining $H_{\rm c2}$ and $H_{\rm irr}$:}  a) Relationship between ${\rm ln}\rho_0$ and $U_0$ for $H\parallel c$ and $\parallel ab$. b) In-field resistivity traces $\rho(T)$ measured in static field up to 35\,T for $H\parallel c$ and c) $H\parallel ab$. For determining $H_{\rm c2}$, a constant resistivity criterion for which the normal state resistivity ($\rho_{\rm n}$) at 28.5\,K is reduced to 90\% ($\rho_{\rm n,0.9}$) is shown as the dotted line. d) and e) The corresponding semi-logarithmic plots. For determining $H_{\rm irr}$, a resistivity criterion of $\rho_{\rm c}=E_{\rm c}/J_{\rm c,100}=1.0\time10^{-8}\,{\rm \Omega cm}$ is shown. Here $E_{\rm c}$ is the electric field criterion ($1\,{\rm \mu V/cm}$) for determining $J_{\rm c}$ from $E-J$ measurements and $J_{\rm c,100}$ is the criterion ($100\,{\rm A/cm^2}$) for determining $H_{\rm irr}$ from $J_{\rm c}-H$ measurements, respectively.}
\label{fig:figures2}
\end{figure}

\clearpage
\subsection{Linear presentation of $E-J$ curves at 4.2\,K}
\begin{figure}[h]
	\centering
		\includegraphics[width=12cm]{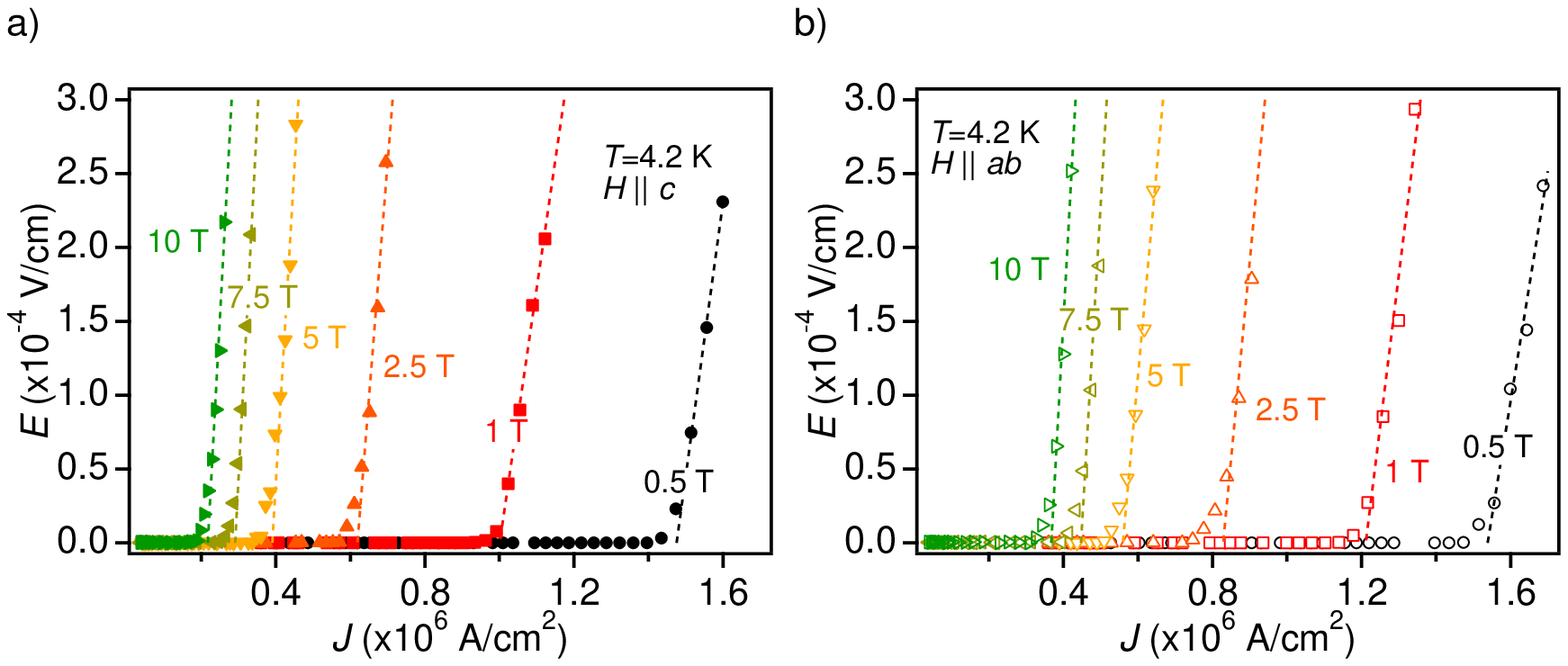}
		\vspace{4cm}
		\caption{{\bf $E-J$ curves:} Linear presentation of the $E-J$ curves shown in Fig.\,4 up to 10\,T for a) $H\parallel c$ and b) $H\parallel ab$.}
\label{fig:figures3}
\end{figure}

\end{document}